**Natural selection from the perspective of mathematical and physical laws**

Running title: Adaptation beyond mathematical and physical laws

Kai Xu

Fisheries College, Jimei University, Xiamen, 361021, China

Correspondence: kaixu@jmu.edu.cn

## Abstract

The theory of evolution by natural selection cannot be used to evaluate the truth value of the following proposition: Through evolution, there exists at least one species that can adapt to any one given environment. To address this issue, this study attempted to define natural selection from the perspective of mathematical and physical laws. This study roughly classified biological activities into molecular, cellular, individual, ecological, and biogeochemical (atomic) levels according to scale and complexity, and selected typical phenomena from each level to analyze the relationship between adaptive evolution and several laws of mathematics and physics. Then, we proposed that natural selection favors heritable variations that allows organisms to better use and/or hide the laws in a certain environment. Reproductive advantage is by far the most obvious consequence of natural selection. Moreover, adaptive evolution can lead to the emergence of laws that ensures that each law only controls some properties of organisms. This study found that organisms significantly influence themselves in all five levels of biological activities, and the whole biosphere can be considered as a huge and evolution-driven feedback loop. Organisms can carry more laws than non-living matter, but the carrying capacity is limited. Therefore, this study's findings suggest that adaptive evolution makes organisms subject to more laws of mathematics and physics until the highest carrying capacity is reached.

## 1. The upper limits of adaptive evolution

To explore the formation mechanisms behind the rich and colorful array of species, Charles Darwin and D'Arcy Thompson focused their attention on the biological changes and laws of mathematics and physics, respectively. Darwin's nature selection theory is one of the mechanisms of evolution, but it is the only mechanism that can lead to adaptive evolution (Urry et al., 2020). Natural selection is a filter that favors heritable variations that provide reproductive advantage in a certain environment. However, this theory still faces some conceptual difficulties, such as lack of fundamental understanding on the existence of biological phenomena and difficulty in quantification (Goldenfeld and Woese, 2011). Thompson suggested that the growth and morphology of organisms must obey the laws of mathematics and physics (Thompson, 1942). Both Darwin's and Thompson's opinions have received abundant support, which means that neither opinion is comprehensive. In addition, the emergence of orders or laws is a common phenomenon for both living and non-living matter (Anderson, 1972), but the relationship between adaptive evolution and the laws of mathematics and physics remains unclear.

Natural selection theory cannot be used to evaluate the truth value of the following proposition: Through evolution, there exists at least one species that can adapt to any one given environment. Thus, we cannot infer the upper limits of adaptive evolution through this theory. To address this issue, we attempted to establish a general theoretical framework by integrating the opinions of Darwin and Thompson. Recent studies have shown that the two dimensional (2D) biological polygon network obeys five mathematical laws but uses oriented cell division to hide the effect of one of these laws on cell size (Xu, 2019a; Xu, 2019b; Xu, 2021; Xu et al., 2017). Thus, the fact that an organism obeys a certain law does not mean that the law must be manifested in the characteristics of organisms. Inspired by this notion, we attempted to define natural selection from the perspective of mathematical and physical laws.

## 2. Five levels of biological activities

From atomic to geochemical scales, interactions between the environment and organisms are shaping the entire Earth's biosphere, and we humans play an increasingly important role in these interactions. Thus, understanding adaptive evolution requires seeing the big picture. In this study, we roughly classified biological activities into molecular, cellular, individual, ecological, and biogeochemical (atomic) levels according to scales and complexity (Fig. 1), and selected typical phenomena from each level to analyze the relationship between adaptive evolution and several laws of mathematics and physics. These laws or orders not only emerged in evolution from simple to complex, but also in the development of multi-celled organisms (Fig. 2). Although the biological mechanisms behind the adaptive phenomena need further in-depth study, the associated mathematical and/or physical laws have been well investigated in previous studies. Below, the relationship between biological phenomena and associated laws is discussed.

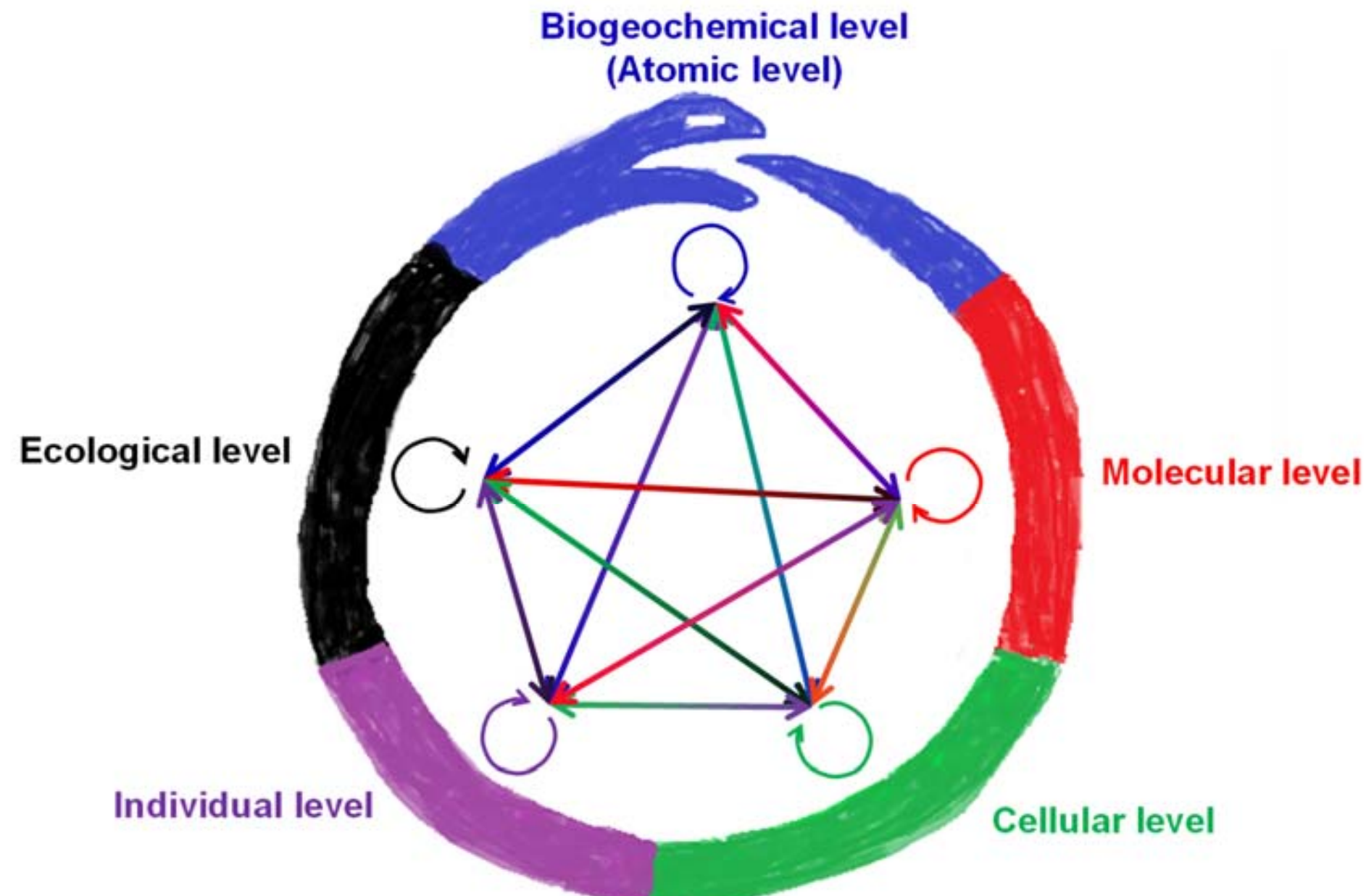

Figure 1 Feedback is everywhere in all five levels of biological activities. From atomic to biogeochemical levels, biological activities are increasing in spatial-temporal scale and complexity. Biogeochemistry is concerned with the collective effects of biological activities at the atomic level.

### 2.1 Molecular level: biochemical reactions

The metabolism is a self-reproducing and self-accelerating network of biochemical reactions. In theory, all chemical reactions are reversible; that is, products in one direction become reactants in the opposite direction. Under given conditions, the direction in which a certain reaction is energetically possible is not determined by organism itself but must obey the laws of thermodynamics (Urry et al., 2020). Whether the reaction will actually occur or not and the reaction rate are not determined by the laws of thermodynamics but can be largely controlled by organism itself. Two methods, spatial-temporal separation and acceleration, are used to gradually and effectively produce desired chemicals.

Spatial-temporal separation is used to place the right amount of chemicals at the right time and position. The formation of the cell membrane is a sign of the emergence of primitive organism, which can be regarded as the oldest version of spatial-temporal separation. As the physical boundary of the cell, membranes not only prevent the free diffusion of chemicals across the membrane but also provide a specific and stable physical, chemical, and biological environment. For eukaryotes, the cell is further compartmentalized to separate the structures with different functions, and compartmentalized structure, such as vacuoles, can switch their functions under different situations (Urry et al., 2020). In addition, there are serious conflicts between many key metabolic pathways, for example, respiration and the toxic effects of oxygen on biological macromolecules and related metabolic pathways, competition between photorespiration and photosynthesis in photoautotrophs, and the paradox between water loss and $CO_2$ uptake through stomata in crassulacean acid metabolism plants (Benzie, 2000; Falkowski and Raven, 2013; Li, 2016; Urry et al., 2020). These metabolic conflicts are partly eliminated by spatial and/or temporal separations.

Spatial-temporal separation only ensure that the desired products will be produced by metabolism, while the production rate is speed up by enzymes. The enzymes are macromolecules that act as catalysts, accelerating chemical reaction by lowering the energy barrier and showing the same effect on the rate of reversible reactions in both directions (Urry et al., 2020). Generally, enzymes can increase the rate of chemical reactions by at least five orders of magnitude (Nelson and Cox, 2017), and then the turnover rate of energy and matter is accelerated by enzymes. This is an important advantage, and it is believed that the chemical reactions in all living things are catalyzed by enzymes (Urry et al., 2020). Thus, enzymes accelerate the cycling of energy and matter in the biosphere. For example, photosynthesis, in which enzymes are involved, plays an important role in the global carbon cycle and can potentially be used to maintain atmospheric $CO_2$ concentration (Falkowski and Raven, 2013; Li, 2016).

In summary, spatial-temporal separation and acceleration ensure the production of biological macromolecules, and are the metabolic mechanisms that use the laws of thermodynamics.

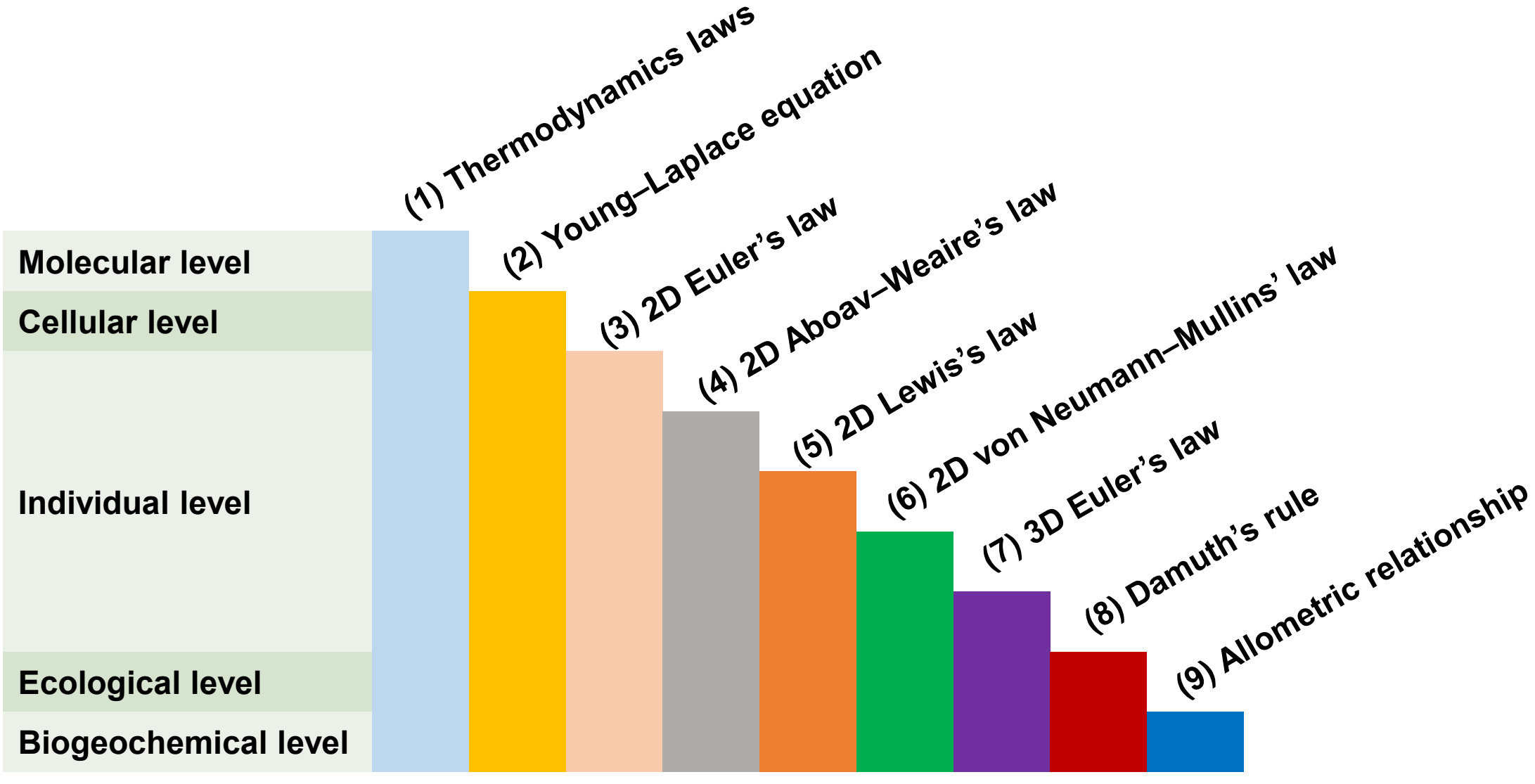

Figure 2 Typical laws that emerge at different levels of biological activity. (1) Biochemical reactions obey the laws of thermodynamics and are accelerated by enzymes. (2) Protoplasts tend to be spherical because they are governed by the Young–Laplace equation, but most cells are non-spherical under real conditions. (3) Epithelium is a 2D space-filling structure composed of polygonal cells, and its topological features can be described by Euler's law. (4) The neighboring relationship between cells can be described by Aboav–Weaire's law. (5) According to Lewis's law, the size of an $n$-edged cell tends to increase with $n$. (6) Division is used to hide the effect of the von Neumann–Mullins law on cell size. (7) The topological features of an early embryo, a three-dimensional (3D) space-filling structure composed of polyhedral cells, can be described by 3D Euler's law. (8) The abundance of wild species follows Damuth's rule, while the biomass of humans and livestock is unusually high among mammals. (9) Both the metabolic rate of organisms and the development of human civilization can be predicted by the allometric

relationship, but the allometric exponents are significant different.

## 2.2 Cellular level: shape of protoplast

Organisms are open systems that exchanges energy and matter with the environment. More importantly, these exchanges take place under the control of organism through complex structures and metabolism (Urry et al., 2020). It is believed that single-celled organisms that are fully enclosed by membranes are the simplest forms of life known to be able to survive independently. The cell membrane forms a selective barrier that controls not only which chemical can cross the membrane but also the rate of crossing (Urry et al., 2020). Therefore, it is widely accepted that the most basic structure of a biological open system is the cell membrane, and the shape of the membrane plays an important role in metabolism. Cell shape is the manifestation of mechanical properties as well as a dynamic consequence of the internal and external forces of the cell. In addition, there are feedback regulations between metabolism, cell shape, and mechanical properties (Si et al., 2015; Ursell et al., 2014).

Surface tension tends to make the bubble spherical according to the Young–Laplace equation. For a given volume, the surface area of spheres is minimal, which not only saves the cost on the boundary but also minimizes the surface energy. For these reasons, the protoplasts of onions become spherical once the rigid cell walls are removed (Taiz and Zeiger, 2010). Surface tension also contributes to the formation of spherical lipid droplets covered by a single layer of membrane (Ben M'barek et al., 2017). The simplest artificial single-celled organism, JCVI-syn3.0, contains less than 500 genes and is also almost spherical in shape (Pelletier et al., 2021). However, the shape of single-celled organisms is highly variable. For example, several dominant phytoplankton species, including diatoms and dinoflagellates, are non-spherical (Sun and Liu, 2003; Thompson, 1942). In this case, the Young–Laplace equation is no longer the main force controlling cell shape.

We can derive the mechanisms of shape transformation simply from the following first principle: The first step must be to hide the influence of surface tension on the cell membrane, which is simple for today's organisms but a huge step forward for the weak primitive membrane-bearing organism. The cytoskeleton is a network of protein polymers that govern the shape and mechanical stability of the cell (Ingber et al., 2014). All three domains of organism, including Bacteria, Archaea, and Eukarya, use cytoskeletons to help maintain cell shape (Wickstead and Gull, 2011). These studies suggested that the cytoskeleton is used to hide the effect of surface tension on the cell shape.

### 2.3 Individual level: space filling

According to the geometric complexity of cell arrangement in life cycle, this study divided organisms into four categories: zero- (0D), one- (1D), two- (2D), and three- (3D) dimensional organisms. A 0D organism is a single cell. The remaining three are multi-celled organisms and can be considered as a set of cells arranged into 1D, 2D, and 3D structures. From a life-cycle perspective, high-dimensional organisms generally develop from a single cell and may experiences the 1D and/or 2D stage before reaching the highest complexity. Besides, parts of a high-dimensional organism can be regarded as low-dimensional structures. A well-known example is the epithelium, a 2D membranous structure consisting of one or more layers of cells that cover most internal and external surfaces of a multi-celled organism and its organs.

Just like the cell membrane of a cell, the epithelium separates the organism or the organs from the external environment, controls the exchanges of energy and matter, protects internal cells, and is the first and main barrier preventing microbial invasion. Thus, the epithelial cells need to be in close contact with each other, leaving no gaps. Each epithelial cell can be simplified as a convex polygon, where every three polygonal cells meet at a vertex. Then, the epithelium can be considered as a 2D polygonal network. Such trivalent 2D structures are found everywhere in nature, from the atomic to astronomic scales (Weaire and Rivier, 1984; Xu, 2021). The geometric and topological dynamics of non-living trivalent 2D structures can be described by four laws: Euler's law, Lewis's law, Aboav–Weaire's law, and the von Neumann–Mullins law (Stavans, 1993; Weaire and Rivier, 1984; Xu, 2021).

As a master rule, Euler's law not only describes the topological relationships between the number of cells, edges, and vertices of trivalent 2D structures, but also applies to any kind of 2D space-filling structures. According to Lewis's law, the cell size increases with the number of edges. Aboav–Weaire's law tell us that cells with few edges tend to neighbor cells with many edges, and verse visa. The change rate of cell size can be described by the von Neumann–Mullins law. These laws show a deep connection between the global and local structures and are related to the conserved distribution of the number of cell edges (Xu, 2021).

The 2D soap froths are one of the most intensively studied non-living trivalent 2D structures, and therefore we compared them with living 2D structures. Based on previous studies, we concluded that as the complexity of cell arrangement in biological structures increases, the above four laws are hidden one by one and may manifest in new aspects of biological activities. Cell size variation in 2D froth structures depends on the number of cell edges, whereas that in living 2D structures is independent of the number of edges (Glazier and Weaire, 1992; Xu, 2021). This difference occurs because the cell size of living 2D structures is mainly controlled by division, a topological process, rather than by the von Neumann–Mullins law (Xu, 2021). A recent simulation study suggested that relaxation drives the geometrical and topological dynamics of trivalent 2D structures, and the von Neumann–Mullins law is one of the consequences of relaxation (Xu, 2021). Another consequence of relaxation is that the intracellular angles are concentrated around 120°, which is consistent with the observations of 2D soap froths and living 2D structures *Pyropia* thalli (Stavans and Glazier, 1989; Xu, 2021; Xu et al., 2017). Interestingly, the relaxation follows only one simple mathematical rule: the central angle between two adjacent vertices of an $n$-edged regular polygon is $2\pi/n$ (Xu, 2021). Thus, the living 2D structure takes advantage of the mathematical rule but hides one of its consequences, namely the von Neumann–Mullins law.

Nearly a century ago, the epidermal pavement cells of plant leaves already attracted much attention owing to their sinuous cell wall (Thompson, 1942). This phenomena is a good example of understanding the relationship between adaptive evolution and laws of mathematics and physics. In the initial stage of development, walled cells can be simplified as convex polygons. Later, cells stop dividing and start expanding and can end up with a size more than 100 times the original (Sapala et al., 2018). This will increase the internal pressure of the epidermal cell, leading to high mechanical stress on the cell wall. To relieve the mechanical stress, the cell wall is bent by the cytoskeleton (Smith and Oppenheimer, 2005). Theoretically, the deformation of the cell wall does not influence the edge number of cells; then neighbor relationship still follows the Aboav–Weaire's law. If the changes in cell size and the deformation of cell wall are uniform, then Lewis's law will be preserved. However, it has been noticed that the pavement cells of *Arabidopsis* leaves do not follow the Aboav–Weaire's law and the Lewis's law (Carter et al., 2017). This deviation may have occurred because of the asynchronous development of cells, as small convex cells can usually be observed coexisting with large pavement cells on the surface of the leaves.

Oriented cell division is essential for morphogenesis of multi-celled organisms, while cell proliferation and tissue stress are linked by mechanical feedback loops (Godard and Heisenberg, 2019). For 1D and 2D biological structures, the direction of cell division is perpendicular to the direction of cell proliferation. As for 3D biological structure filling with polyhedral cells, the division can happen at any direction. The 2D version of Euler's law is mathematically a specific case of the 3D version, but in biology, they need to be executed by different mechanisms. Moreover, in the 3D condition, Euler's law is not sufficient in identifying whether a polyhedron is sealed or not (Grünbaum and Motzkin, 1963; Thompson, 1942). This is a serious problem for the study of 3D space filling. We should learn from nature, as some organisms evolved exquisite and robust mechanisms to follow the 3D Euler's law, for example, the coccolithophore *Emiliania huxleyi*, the radiolarias *Didymocyrtis*, and *Pantanellium* (Xu et al., 2018; Yoshino et al., 2019; Yoshino et al., 2015; Young et al., 2003).

The early embryo can be considered as a 3D space-filling structure composed of convex polyhedral cells, which means it obeys 3D Euler's law. The most attractive cells in animals are neurons, which are the basis of consciousness. The unique shape of neurons makes the human nervous system not a 3D structure filled with polyhedral cells, but essentially a network of countless connected neurons. In this case, Euler's law can be used to describe the topological properties of neural networks (Santos et al., 2019).

Human fusiform HeLa cells and astrocytes can be converted into functional neurons, and related technologies are expected to be used for the treatment of nervous system diseases (Qian et al., 2020; Xue et al., 2013). These analyses and experimental studies suggested that the evolution of animal consciousness is based on the complex application of the laws of mathematics and physics. In addition to the Euler characteristic, another topological invariant, the Betti number, has been used to analyze the topological phase transition of the brain neural network (Santos et al., 2019).

### 2.4 Ecological level: size and abundance

About 80% of the Earth's biomass is plants, mainly terrestrial trees (Bar-On et al., 2018). Bacteria are the second major biomass component, accounting for about 15% of global biomass. Given the huge size difference between bacteria and trees, it is not surprising that the number of bacteria is estimated to be more than $10^{15}$ times that of trees (Bar-On et al., 2018). This is consistent with a previous study that used a power law known as Damuth's rule to describe the relationship between the size and abundance of photoautotrophs, including marine phytoplankton and terrestrial plants (Belgrano et al., 2002). According to this law, the abundance of organisms decreases with increasing size. Damuth's rule has also been found in many other organisms, such as animals and bacteria, although the scaling exponent may differ owing to taxonomic and/or environmental differences (Gjoni and Glazier, 2020).

Based on Damuth's rule, if the variation of the Earth's surface environment is neglectable, then the distribution of different-sized organisms should remain relatively stable. However, the biomass of mammals at present is about four times that before human civilization (Bar-On et al., 2018). There are two reasons for this: the biomass of wild mammals has decreased by about 80% owing to the development of human civilization; at present, the biomass of both humans and livestock is more than 10 times the biomass of wild mammals (Bar-On et al., 2018). The increase in mammalian biomass is not due to the fourfold increase of resources on the Earth's surface, but rather due to the increasing ability of humans to obtain and utilize resources. These data suggested that humans do not obey Damuth's rule when considering species abundance.

Similar power-law relations, such as Zipf's law and the Pareto principle, are reflected in many aspects of human civilization and nature, including the distribution of firms, wealth, income, words, cities, species abundance, sand, and meteorites (Lü et al., 2010; Su, 2018; West, 2017). According to Zipf's law, if words are ranked in descending order of frequency, frequency is linearly related to the inverse of the rank (Zipf, 1949). A recent study analyzed eight datasets of wild species from 11 taxonomic groups and found that species abundance followed Zipf's law (Su, 2018). In addition, Zipf's law could also be used to describe the distribution of about 30,000 cities, and the exponent did not change over the years 1992, 2001, and 2010 (Jiang et al., 2015).

### 2.5 Biogeochemical (Atomic) level: size and metabolic rate

Under the action of physical forces, fundamental particles form chemical elements, which are then organized to form stars (Li, 2016; Schlesinger and Bernhardt, 2020; Viola, 1990). Additionally, the motion and evolution of stars and various matter are also controlled by the physical forces. As one of the four fundamental forces of nature, gravity is one of the main drivers that determine the distribution of elements on the earth, especially as it enriches five of the six essential elements, namely carbon (C), hydrogen (H), nitrogen (N), oxygen (O), and sulfur (S), on the earth's surface (Li, 2016; Schlesinger and Bernhardt, 2020). This is the atomic basis for the emergence and continuation of organisms.

All living organisms need to exchange energy and chemical matter with the external environment. Six essential elements, C, H, N, O, phosphorus (P), and S, account for more than 95% of the biomass on the Earth (Li, 2016; Schlesinger and Bernhardt, 2020). All organisms, including microorganisms, plants, and animals, have a similar elemental composition. Therefore, this elemental preference is a general phenomenon, indicating that organisms compete mainly for the above six elements. Elemental preference is also reflected in the carbon isotopes. Owing to minor chemical and physical differences, lighter carbon isotopes are much more likely to be fixed by photosynthesis, which results in lower $^{13}C/^{12}C$ ratios in photoautotrophs than in the atmosphere (Craig, 1953; Farquhar et al., 1989). Fossil fuels, the remains of ancient organisms (mainly plants) from about 300 million years ago, have a $^{13}C/^{12}C$ ratio similar to those of modern plants (Craig, 1953). This proves that the elemental preference of organisms has been present for about 300 million years. The collective effects of biological activities at the atomic level are becoming an increasingly important driver of the large-scale cycling of related elements. This phenomenon is a hot topic for biogeochemical research. In addition, organisms can influence themselves in all five levels of biological activities through global and local cycles of elements. Thus, the biosphere can be considered as a huge feedback loop.

Julian Huxley and Max Kleiber summarized an empirical power law, the allometric relationship, to describe the relationships between individual size and metabolic rate (Brown et al., 2004; Huxley, 1932; Kleiber, 1932). This law has been reexamined and confirmed across a mass range of more than 25 orders of magnitude, and the exponent is in the range of 0–1 (Brown et al., 2004; West, 2017). According to this law, the metabolic rate per individual tends to increase with increasing biomass; but the metabolic rate normalized by biomass tends to decrease, indicating a decrease in metabolic efficiency. The allometric relationship of metabolism can be used to explain why both marine phytoplankton and terrestrial plants contribute about 50% of global net primary production, although the former accounts for only about 0.2% of the latter's biomass (Allen et al., 2005; Field et al., 1998).

Recently, the allometric relationship has also been observed in many aspects of human civilization; for example, economic growth rates, crime rates, patent rights, and even walking speed all scale with the population of a city (West, 2017). When this law is applied to quantify the development of human civilization, the exponent becomes larger than 1, indicating that the efficiency of human society increases with size. This explains why over the past several thousand years, more people have moved to cities, and cities have grown larger. The allometric exponent suggests that human civilization is accelerating. A landmark event occurred in 2020: human-made mass surpassed all living biomass (Elhacham et al., 2020). This will have a significant impact on global biogeochemistry and human civilization, but we still know little about it. Interestingly, based on the above studies (Brown et al., 2004; West, 2017), we know that although both human metabolism and human civilization obey the allometric relationship, the exponents are significantly different.

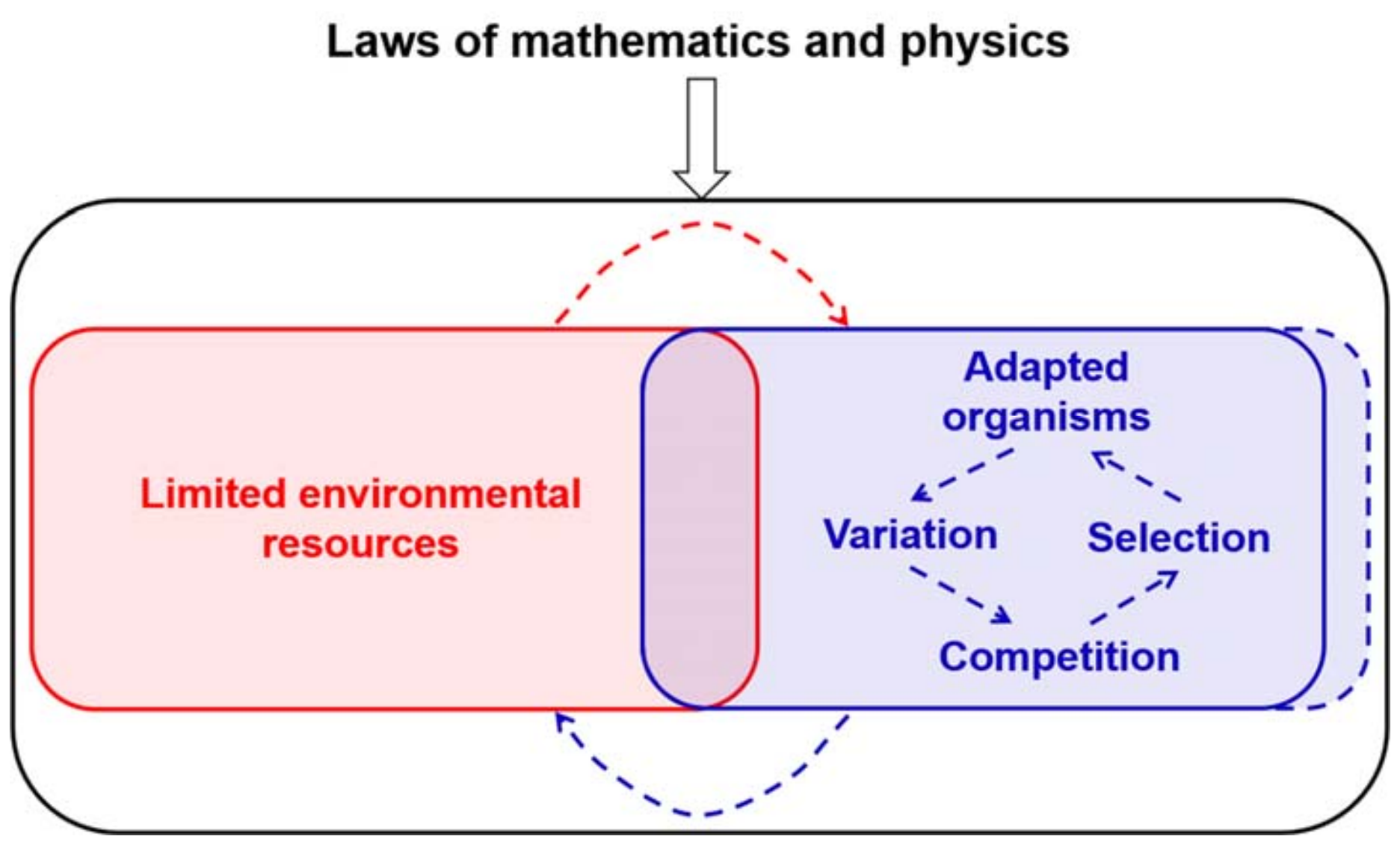

Figure 3 Organisms and the environment form an adaptive evolution-driven feedback

system. Both organisms and non-living things are subject to specific laws of mathematics and physics (including some laws in common), and each law controls only some of their properties. Adaptive evolution is the basis for the emergence of laws.

## 3 Conclusion

Natural selection theory cannot be used to infer the upper limits of adaptive evolution. The key to address this issue is to define natural selection from the perspective of mathematical and physical laws. Thus, this study analyzed typical biological phenomena from the molecular to the biogeochemical (atomic) levels and proposed the concept that natural selection favors heritable variations that allows organisms to better use and/or hide the laws in a certain environment. By far the most obvious consequence of natural selection is the reproductive advantage it can provide.

This study also established a feedback model to explain the relationships between laws of mathematics and physics, the environment, and organisms (Fig. 3). The key points of our model are: 1. Both living and non-living matters are subject to specific laws of mathematics and physics, and each law controls only some of their properties. 2. Adaptation beyond laws rather than simply following them, and this is one of the major differences between living and non-living matter. 3. Adaptive evolution is the basis for the emergence of orders or laws. 4. Adaptive evolution drives the feedback system. Feedback regulation can be found everywhere in all five levels of biological activities, and the whole biosphere can be considered as a huge feedback loop (Figs. 1 & 3). The accelerated development of human civilization (West, 2017) indicates that the feedback loop is running faster.

The other significant difference between living and non-living things comes from the fact that organisms can carry more laws and that the laws are combined and presented in more complex ways (Fig. 2). However, some environmental conditions and associated laws shape organisms in an exclusive way, for example, to adapt to flying, running and swimming, animal bodies have evolved distinctly different geometric and physical features. This suggests that the carrying capacity is not limited. Therefore, adaptive evolution makes organisms subject to more laws until the highest carrying capacity is reached.

**Funding**

This work was supported by the National Key Research and Development Program of China (2018YFD0900702).

**Conflicts of interest**

The author reports no potential conflicts of interest.